\documentclass[aps,pre,reprint,superscriptaddress,longbibliography]{revtex4-1}
\usepackage{graphicx} 

\usepackage{float}       
\usepackage{cancel}

\usepackage{color}
\usepackage{amssymb,amsmath}

\begin{document}

\title{Local analysis of the history dependence in tetrahedra packings}

\author{N. Nirmal Thyagu}
\affiliation{Max Planck Institute for Dynamics and Self-Organization (MPIDS), 37077 Goettingen, Germany}
\affiliation{Division of Physics, School of Advanced Sciences, VIT Chennai, Chennai - 600127, India}

\author{Max Neudecker}
\affiliation{Max Planck Institute for Dynamics and Self-Organization (MPIDS), 37077 Goettingen, Germany}

\author{Simon Weis}
\affiliation{Institut f\"ur Theoretische Physik I, Friedrich-Alexander-Universit\"at, 91058 Erlangen, Germany}

\author{Fabian M. Schaller}
\affiliation{Institut f\"ur Theoretische Physik I, Friedrich-Alexander-Universit\"at, 91058 Erlangen, Germany}
\affiliation{Institute of Stochastics, Karlsruhe Institute of Technology, 76049 Karlsruhe, Germany}

\author{Matthias Schr\"{o}ter}
\email[]{matthias.schroeter@ds.mpg.de}
\affiliation{Max Planck Institute for Dynamics and Self-Organization (MPIDS), 37077 Goettingen, Germany}
\affiliation{Institute for Multiscale Simulation, Friedrich-Alexander-Universit\"at, 91052 Erlangen, Germany}

\date{\today}

\begin{abstract}
The mechanical properties of a granular sample depend frequently on the way the packing was prepared.
However, is not well understood which properties of the  packing store this information. 
Here we present an  X-ray tomography study  of three pairs of tetrahedra packings 
prepared with three different tapping protocols. 
The packings in each pair differs in the number of mechanical constraints  $C$ 
imposed on the particles by their contacts,
while their bulk volume fraction $\phi_\textnormal{global}$ is approximately the same.
We decompose $C$ into the  contributions of the three different contact types possible 
between tetrahedra -- face-to-face (F2F), edge-to-face (E2F), and point contacts -- 
which each fix a different amount of constraints.  
We then perform a local analysis of the contact distribution by
grouping the particles together according to their individual
volume fraction $\phi_{local}$ computed from a Voronoi tessellation.
We find that in samples which have been tapped sufficiently long
the number of F2F contacts becomes an universal function of  $\phi_\textnormal{local}$.
In contrast the number of E2F and point contacts varies with the applied tapping protocol.
Moreover, we find that the anisotropy  of the shape of the Voronoi cells  depends on the tapping protocol.
This behavior differs from spheres and ellipsoids and posses a significant constraint for any
mean-field approach to tetrahedra packings.
\end{abstract}


\maketitle

\section{History dependence in granular matter}
A number of experiments have shown that the mechanical 
properties of apparently identical granular samples will differ 
depending on how they were prepared.
Examples include the volume response to shear \cite{radjai:04}, 
the force moment tensor in tapped packings \cite{pugnaloni:10,pugnaloni:11,ardanza-trevijano:14},
the increase of pressure with depth in a granular column \cite{vanel:99,wambaugh:10,back:11,perge:12}, 
or the pressure distribution \cite{vanel:99_pile} below a sandpile. 
Another type of history dependent behavior, 
which is sometimes referred to as memory effect, can be seen if a sample is compactified, e.g.~by shearing or tapping
with a driving strength  $s_i$, to a specific volume fraction 
$\phi_\textnormal{global}^0$. 
If  the driving strength is then changed to some new value
$s_0$, it is found that the subsequent evolution of $\phi_\textnormal{global}$ differs for different values of  
$s_i$ \cite{josserand:00,nicolas:00}; even though all samples now start
at  $\phi_\textnormal{global}^0$ and are driven with the same strength $s_0$.
A similar effect was found in granular gases \cite{prados:14}.

In some sense the term history dependence simply expresses the fact that $\phi_\textnormal{global}$ alone does not provide a complete
description of the state of the packing. Given that the mechanical properties of the sample originate from
forces transmitted between particles at their contacts it seems natural to extend the description by 
including  information on
a) the number and spatial structure of the contacts, and 
b) the distribution of contact forces.
Moreover we might want to include,  for reasons shown below, 
c) properties of the Set Voronoi cells
surrounding each particle. 
(The Set Voronoi cell of a given particle comprises all points which are closer to the surface of the particle
than to the surface of any other particle \cite{medvedev:94,schaller:13,weis:17}.) 

Option a) has interesting theoretical consequences.
This paper studies pairs of hard, frictional tetrahedra packings which have approximately the same 
$\phi_\textnormal{global}$ but differ in their contact numbers $Z$. Such behavior is clearly beyond the reach
of the Jamming paradigm which has been developed for soft, frictionless spheres \cite{liu:10,vhecke:10} and 
where $Z$ is considered to be a function of $\phi_\textnormal{global}$ alone. 

The reason for this failure is not even primarily the shape of the particles: 
In the Jamming paradigm both $\phi_\textnormal{global}$ and  $Z$ are controlled simultaneously by the compression
of the soft particles. In contrast, granular particles change their contact number by changing 
their local geometry, not by compression. 
For tapped spheres and ellipsoids
\footnote{The limited choices of experimental preparation protocols might have influenced this 
results: it has been shown numerically by Agnolin and Roux (Phys. Rev. E 76,
061302, 2007) that even for spheres the value of $Z$ can differ for identical  $\phi_\textnormal{global}$.}, 
$Z$ is controlled  by the local volume fraction
$\phi_\textnormal{local}$  which is computed for each individual particles by 
dividing its volume by the volume of its Voronoi cell
\cite{schaller:15}.
This behavior is in good agreement with statistical mechanics approach \cite{song:08,baule:13,baule:14}, 
where $Z$ is computed from a mean-field approximation of $\phi_\textnormal{local}$.

Recently there has been some effort to evolve the Jamming paradigm into a theory which describes 
history dependence \cite{kumar:16,luding:16}; if this approach can be expanded to 
describe also frictional and hard particles it would be very valuable.  

Option b) is directly connected to the fact that 
frictional packings at finite pressures are hyperstatic \cite{silbert:02,zhang:05,shundyak:07,henkes:10,schroeter:17}
i.e.~there exist many possible force configurations which fulfill a set of given boundary conditions \cite{tighe:10}. 
Measuring the distribution of contact forces experimentally is a hard problem; it has presently 
only been achieved in systems of compressive particles \cite{tsoungui:98,dijksman:17} or photoelastic discs \cite{daniels:17}.
Experiments with the latter method find a clear dependence on the preparation protocol \cite{majmudar:05} . 

Option c) might initially not look too promising: 
at least for spheres neither the distribution of the volumes of the Voronoi cells \cite{aste:07} 
nor the distribution of their shapes \cite{schroeder-turk:10} shows  
any signature of the preparation conditions. However,
for packings of tetrahedra we find that the preparation history 
does indeed influence the shape and size distribution of the Voronoi volumes.

\section{Packings of tetrahedra}
\label{sec:tetra_pack}
An interesting difference between spheres and tetrahedra 
is that the latter have not only one but four different types of contacts:  
face-to-face (F2F) contacts, 
which are mechanically equivalent to 3 individual  point contacts, 
edge-to-face (E2F) contacts (equivalent to two point contacts) and the vertex-to-face (V2F) 
and edge-to-edge (E2E) contacts which we group here as point contact.

This distinction is important because the different contact types fix each a different number of 
degrees of freedom. The total number of constraints of a particle $C$ is therefore $\sum_i C_i$ where
$i$ goes over the three contact types (F2F, E2F, point). Each $C_i$ equals $m_i Z_i$ where
$Z_i$ is the respective contact number
and $m_i$ the number of constraints per particle fixed by this type of contact. The different 
values of $m_i$ can be understood for frictional contacts as follows: 
All contacts impose 3 translational constraints, E2F contacts add 2 rotational constraints (thus, a total of 5 constraints),
F2F contacts prohibit 3 different rotations (thus, a total of 6 constraints).
As these constraints  are shared between two tetrahedra, we obtain the  
constraint multipliers $m_{F2F}=3.0$, $m_{E2F}=2.5$, and $m_{V2F}=m_{E2E}=1.5$.

Each tetrahedron has 6 degrees of freedom (3 translations and 3 rotations), 
therefore a mechanical stable packing requires
$C$ to be  at least 6, the so-called isostatic contact number.  However, typical packings of frictional
tetrahedra have $C$ values in the range 12 to 18 \cite{neudecker:13} 
and are therefore strongly hyperstatic \cite{zhao:17}. 
Contrary claims of isostaticity \cite{jaoshvili:10} are based on 
the erroneous use of frictionless constraint multipliers $m_i$ while analyzing experimental i.e.~frictional packings. (Due to the absence of tangential forces the frictionless $m_i$ are always smaller than their frictional counterparts.)  

Because of the importance of the constraint number for the mechanical properties, 
we will discuss in the remainder of this paper $C$ rather than $Z$.

Packings of frictionless tetrahedra have recently been an active field of research because of
the existence of  quasi-crystals \cite{haji-akbari:09}, hyperuniformity \cite{jiao:11},
the effect of altered particle shape on packing density \cite{kallus:11,zhao:12,damasceno:12,jin:15b},
the distribution of contact types \cite{smith:10,smith:11,smith:14}, and the behavior of 
binary systems of tetrahedra and octahedra \cite{cadotte:16}. Theoretical approaches
include a mean field ansatz \cite{baule:13,baule:14} and a
local motif analysis \cite{li:13,liu:14,jin:15}.
Moreover there has been a quest for the densest tetrahedra packing \cite{torquato:09,kallus:10,torquato:10}, 
which is  presently believed to be at $\phi_\textnormal{global} \approx 0.8563$ \cite{chen:10}.

Experiments using necessarily frictional tetrahedra explore a much lower range of $\phi_\textnormal{global}$ \cite{baker:10}  than the numerical studies of frictionless tetrahedra described above.
They have been characterized with respect to their
mechanical properties under uniaxial compression \cite{athanassiadis:13,jaeger:15} and shown to have
history dependent values of $C$ and $Z$ for identical values of $\phi_\textnormal{global}$  \cite{neudecker:13}.


\section{Experimental setup}
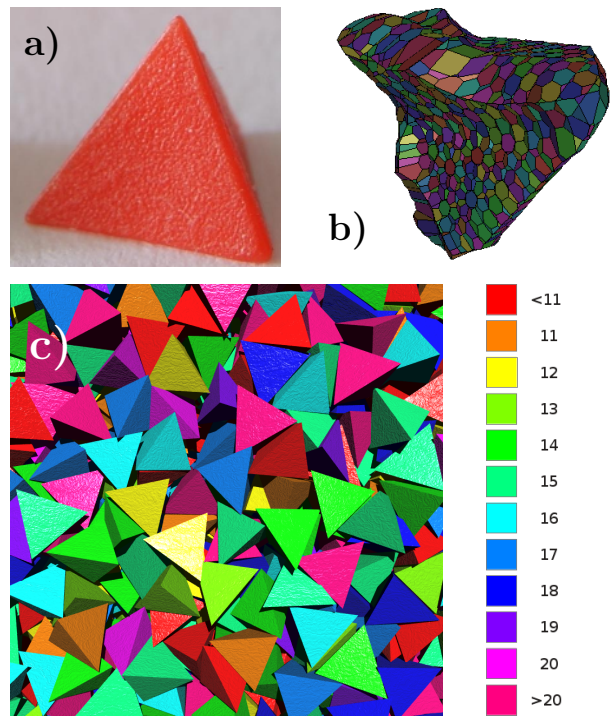
\begin{figure}[t]
\begin{picture}(270,270)
\put(10,170){\includegraphics[width=0.20\textwidth]{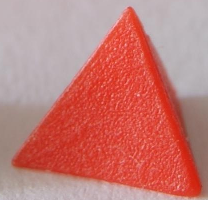}}
\put(130,170){\includegraphics[width=0.22\textwidth]{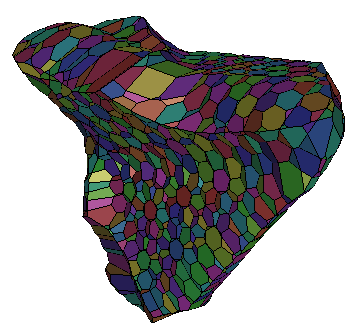}}
\put(10,0){\includegraphics[width=0.32 \textwidth]{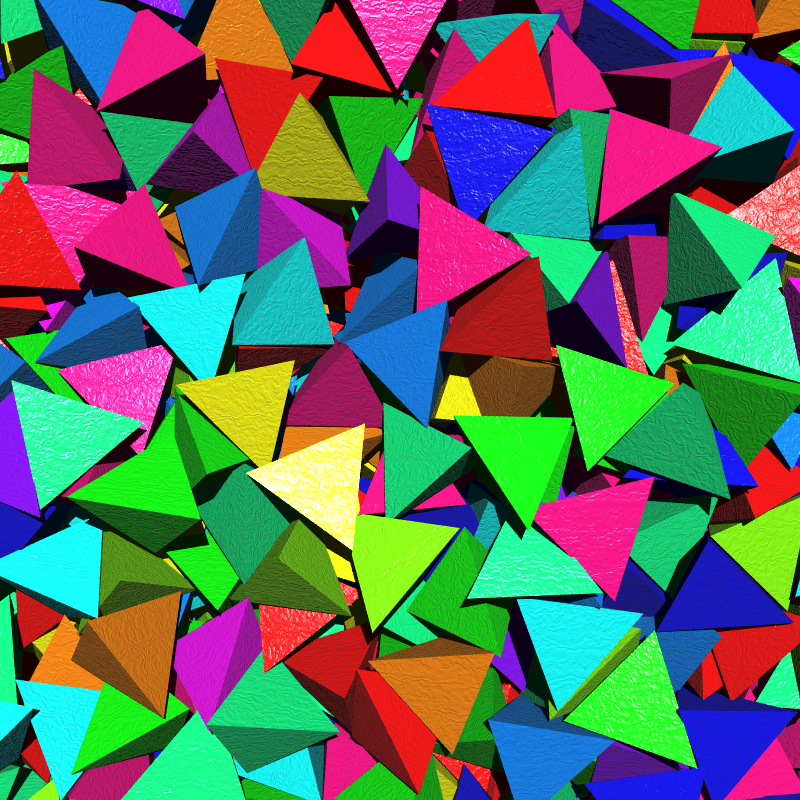}}
\put(190,0){\includegraphics[width=0.062\textwidth]{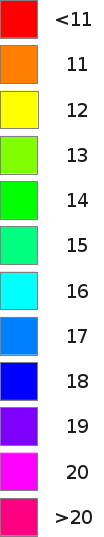}}
\put(15,250) {\Large \bf a)}
\put(130,180) {\Large \bf b)}
\put(17,135) {\LARGE \color{white} \bf c)}
\end{picture}
\caption{Tetrahedra packing.  
{\bf a)} Polypropylene particle made by injection moulding, the edge length is 7mm.
{\bf b)} Set Voronoi cell around a tetrahedron with a local volume fraction of 0.52. The Set Voronoi tessellation 
is computed by spawning 514 point on the surface of each tetrahedron and them merging the Voronoi cells of these
points.
{\bf c)}  Rendering of an interior section of a tetrahedra packing after all particles have been 
identified from an X-ray tomogram.  
Particles are colored according to the number of constraints fixed by their neighbors.  
The global volume fraction is 0.53, the average constraint number is 15.5.
}
\label{fig:rendering}
\end{figure}

Experiments were performed with the same polypropylene tetrahedra as in reference \cite{neudecker:13}.
The particles, figure  \ref{fig:rendering}a shows an example,  
were made by mould casting, their edge length is 7 mm and their coefficient of 
static friction equals 0.8 \cite{note_friction}.

\subsection{Preparing packings}
Packings were prepared by tapping, 
starting from a loose initial configuration which was prepared by 
first filling the particles into a smaller inner cylinder (without bottom) 
and then releasing them into a larger cylindrical container of diameter 10.4 cm
by slowly moving the inner cylinder upwards. Tapping is done with an
electromagnetic shaker (LDS model V555) which is driven by 
a series of individual sinusoidal pulses with a periodic time of 1/3\,s. The separation 
between individual pulses is 0.5\,s, the amplitude of the sinusoidal motion 
is chosen such that the peak acceleration $\Gamma$ corresponds to either 2g, 5 g, or 7g 
where $g$ is the acceleration due to gravity.

To monitor the evolution
of $\phi_\textnormal{global}$ during tapping, height profiles of the packing 
are measured in regular intervals with a laser distance sensor (MicroEpsilon ILD1402) 
mounted on a horizontal translation stage. These readings are then corrected for boundary 
effects by calibrating them with the bulk $\phi_\textnormal{global}$ values measured 
from the tomographic reconstructions.

\begin{figure}[t]
\centering
\includegraphics[width=0.5\textwidth]{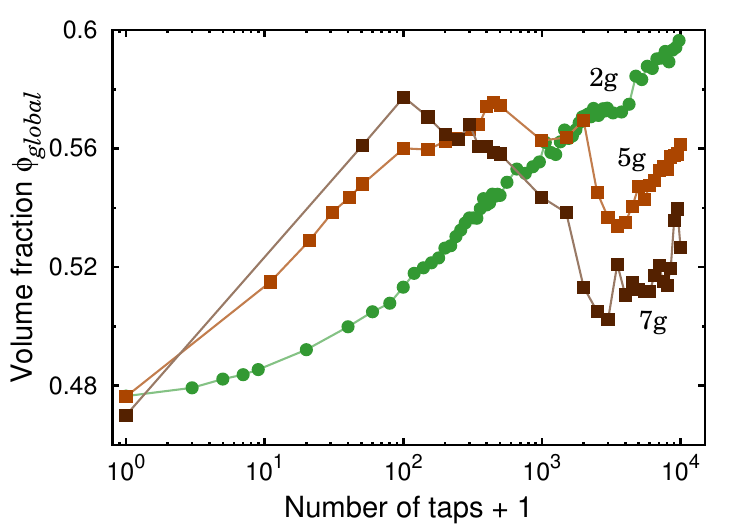}
\caption{Compaction of tetrahedra packings for three different tapping strengths $\Gamma = 2g, 5g$ and $7g$.
}
\label{fig:compaction}
\end{figure}

Figure \ref{fig:compaction} demonstrates the evolution of  $\phi_\textnormal{global}$ with the number of taps.
For a tapping strength $\Gamma$ = 2g there is a monotonic increase in $\phi_\textnormal{global}$.
In contrast, at $\Gamma$ = 5g and 7g the volume fraction first increases, then decreases and finally  increases again. 
In none of the experiments a steady state is reached; in  reference \cite{neudecker:13} it is shown that 
at least 10$^5$ taps are needed at $\Gamma$ = 2g for reaching a plateau in $\phi_\textnormal{global}$.

\subsection{Image analysis}
After the packings have been prepared, 
three-dimensional images of the geometrical arrangement of particles are obtained by
X-ray tomography (Nanotom, GE) with a resolution of 100 $\mu$m per voxel.
Particles are then identified by a two step algorithm \cite{neudecker:13}
involving a cross-correlation with an 
inscribed sphere and a steepest ascend gradient search; the particle detection rates are better than 99.8 \%.
After the positions and orientations of all particles are known, the number and type of contacts is computed 
for each particle. 

As shown in the supplements of reference  \cite{neudecker:13} the detection of the total number of
contacts has an approximate statistical error of $\pm$ 0.2. 
Unfortunately, we can not derive an error estimate
for the algorithm assigning the different contact types. 
We will therefore consider $\pm$ 0.2 $m_i$ as a 
best case approximation of our actual errors; this corresponds to $\pm$ 0.3 for $C_{point}$,  
$\pm$ 0.5 for $C_{E2F}$, and $\pm$ 0.6 for $C_{F2F}$.

Figure \ref{fig:rendering}c displays a rendering of a cross section through the bulk
of a sample where the individual tetrahedra are color coded according to their number of constraints.  
For the further analysis only particles with a center of mass to container wall distance of at least  2 particle side lengths are 
retained. This corresponds to 2457 to 3893 tetrahedra per packing as shown in table \ref{tab:choice}.

\begin{table*}
\caption{Volume fraction $\phi_\textnormal{global}$, Standard deviation of the local volume fraction distribution $\sigma$,
total constraint number $C$, contribution of the three different contact
types (F2F, E2F, point), Voronoi cell isotropy index $<\beta_0^{2,0}>$, and number  $N$  of tetrahedra analyzed.
\label{tab:choice}
}
\begin{ruledtabular}
\begin{tabular}{c  c|  c  c | c c c c | c | c}
name & symbol& $\phi_\textnormal{global}$ & $\sigma$ &$C$ & $C_{F2F}$ & $C_{E2F}$ & $C_{point}$ &  $<\beta_0^{2,0}>$ & $N$ \\
\hline
{\tt 10G2}  & {\large \color{blue}$\circ$}  & 0.482 & 0.051 &14.4 & 0.81 & 5.8 & 7.7 & 0.678 & 3421 \\ 
{\tt 20G2}  &  {\large \color{blue}$\bullet$}   & 0.481 & 0.048 &13.1 & 0.39 & 5.2 & 7.5 & 0.690 & 3425 \\ 
\hline
{\tt 400G2} & \hspace{0.1em} {\color{red}$\square$ }  & 0.531 & 0.048 &15.5 & 0.63 & 6.3 & 8.5 & 0.724 & 3778 \\
{\tt 10000G7}& {\color{red}$\blacksquare$} & 0.526 & 0.060 &13.7 & 0.80 & 5.6 & 7.4 & 0.708 & 2457 \\
\hline
{\tt 1600G2}& {\color{magenta}$\vartriangle$} & 0.545 & 0.052 &16.0 & 1.05 & 6.6 & 8.4 & 0.728 & 3893 \\
{\tt 10000G5}& {\color{magenta}$\blacktriangle$}  & 0.551 & 0.059 &15.0 & 1.09 & 6.1 & 7.8 & 0.718 & 2580 \\ 
\end{tabular}
\end{ruledtabular}
\end{table*} 

The Set-Voronoi cells of the tetrahedra were computed using the program {\tt Pomelo}
\cite{weis:17} which first discretizes the surface of each particle by spawning
 points on it, then performs a standard Voronoi tessellation based on all surface points,
and finally merges all the standard Voronoi cells on the surface of a given particle to obtain
its Set-Voronoi cell. In this work, we resolve each tetrahedra with 514 surface points;
figure \ref{fig:rendering}b gives an example of the resolution of such a Set-Voronoi cell. 
In order to account for the imperfections
of both our particles and the image processing, we modify the position of the surface points in two ways:
a) we replace the four sharp vertices of the tetrahedra with rounded caps. These cap are approximated by spheres with a radius of $\approx$ 5\% of the side length of a tetrahedron.
And b) we shrink the apparent particle size by 1 \% in order to avoid artifacts due to overlaps.
$\phi_\textnormal{local}$ of the individual particles is then computed by dividing the tetrahedra 
volume by the volume of the Set-Voronoi cells.

The shape of the Set-Voronoi cells is analyzed with the help of Minkowski Tensors of rank two, 
computed with the program {\tt Karambola} \cite{schroder-turk:13}. More specifically 
we compute for each Set-Voronoi cell the Minkowski Tensor $W_0^{2,0}$ 
which bears a resemblance to the moment of inertia tensor of the cell. 
The isotropy index $\beta_0^{2,0}$ is then the eigenvalue ratio of the smallest to the largest
eigenvalue of $W_0^{2,0}$. Thus for a perfectly isotropic cell,
$\beta_0^{2,0}$ will equal one; with increasing anisotropy the value
will decrease. 
Averaging over all particles in the core of the sample we obtain a global isotropy index $<\beta_0^{2,0}>$, for which the origin is the centroid of the Set-Voronoi cell.

\section{Results}

\subsection{History dependence of the average constraint number}

\begin{figure}[t]
\centering
\includegraphics[width=0.5\textwidth]{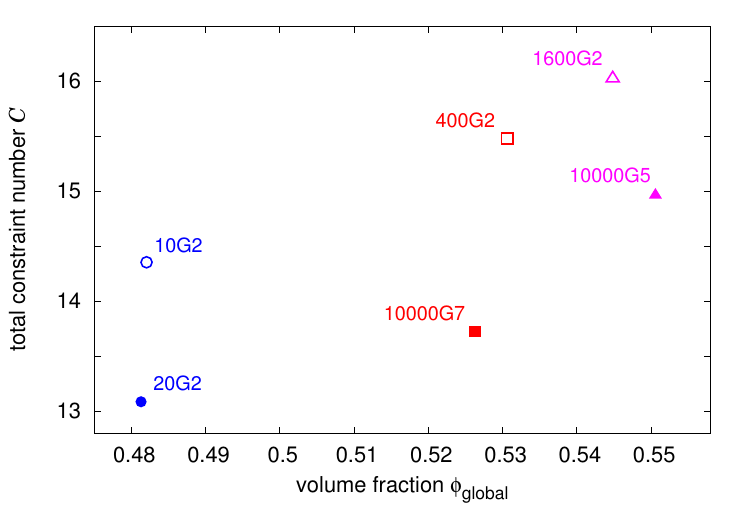}
\caption{History dependent preparation. By choosing the number and strength  of taps 
during preparation, we prepare three pairs of packings with approximately the same 
$\phi_\textnormal{global}$ but different values of $C$.
}
\label{fig:choice}
\end{figure}

 As already shown in \cite{neudecker:13}, the average number of constraints in a 
tetrahedra packing does not only depend on $\phi_\textnormal{global}$ but also on the tapping protocol
used to prepare the sample. 
This history dependence allows us to prepare three pairs of packings with each approximately  identical $\phi_\textnormal{global}$ 
but  different values  in $C$.
These pairs are indicated by open and closed symbols in figure \ref{fig:choice}. 
The naming scheme we will use in the following is based on  $n$G$a$ where $n$ stands of the number of taps applied and 
$a$ corresponds to the acceleration of these taps. I.e.~packing {\tt 1600G2} has been tapped 1600 times at 2 g.

The two denser pairs of packings are formed by  a more constraint packing prepared by tapping
the sample with 2 g and  a  less constraint packing that has been tapped with 5 g and 7 g. Their
names are {\tt 400G2} - {\tt 10000G7}  and {\tt 1600G2} -  {\tt 10000G5}. 
The third pair represents fluctuation of the initial preparation by comparing two different experimental runs both
tapped at 2g:  {\tt 10G2} - {\tt 20G2}. All four of the samples tapped at 2 g were also included in reference \cite{neudecker:13};
the particle positions of all six experiments discussed in this paper can be downloaded from Zenodo \cite{zenodo}. 

Table \ref{tab:choice}  displays  the  $\phi_\textnormal{global}$ and $C$ values of all six experiments discussed in this paper. 
Moreover, the table also shows the contributions of the three different contact types,
$C_{F2F}$, $C_{E2F}$, and  $C_{point}$, towards the total number of constraints.
In all experiments the point contacts contribute with  52 to 57 \% the majority to the total constraint number
while the face-to-face contacts provide only 3 to 7 \% of $C$; even though their constraint multiplier $m$ is twice as
large as  $m$ of the point contacts. We will provide a more detailed analysis of the contribution of the different
contact types in  section \ref{sec:local_constraint_contribution}. 

Table \ref{tab:choice} shows also  that the average isotropy  index $<\beta_0^{2,0}>$ is history-dependent: 
for the two marginally tapped samples the lower constraint packing is composed of more isotropic  
Voronoi cells. For the two pairs tapped at different g values, the more constraint 
packings are more isotropic.  This behavior is in contrast to sphere packings where it was shown in 
\cite{schroeder-turk:10} that $<\beta_0^{2,0}>$ is independent of the preparation; only increasing monotonously with
$\phi_\textnormal{global}$. In section \ref{sec:isotropy_local} we present a local analysis of  $<\beta_0^{2,0}>$.

\subsection{A local view on tetrahedra packings}

\begin{figure}[t]
\includegraphics[width=0.45\textwidth]{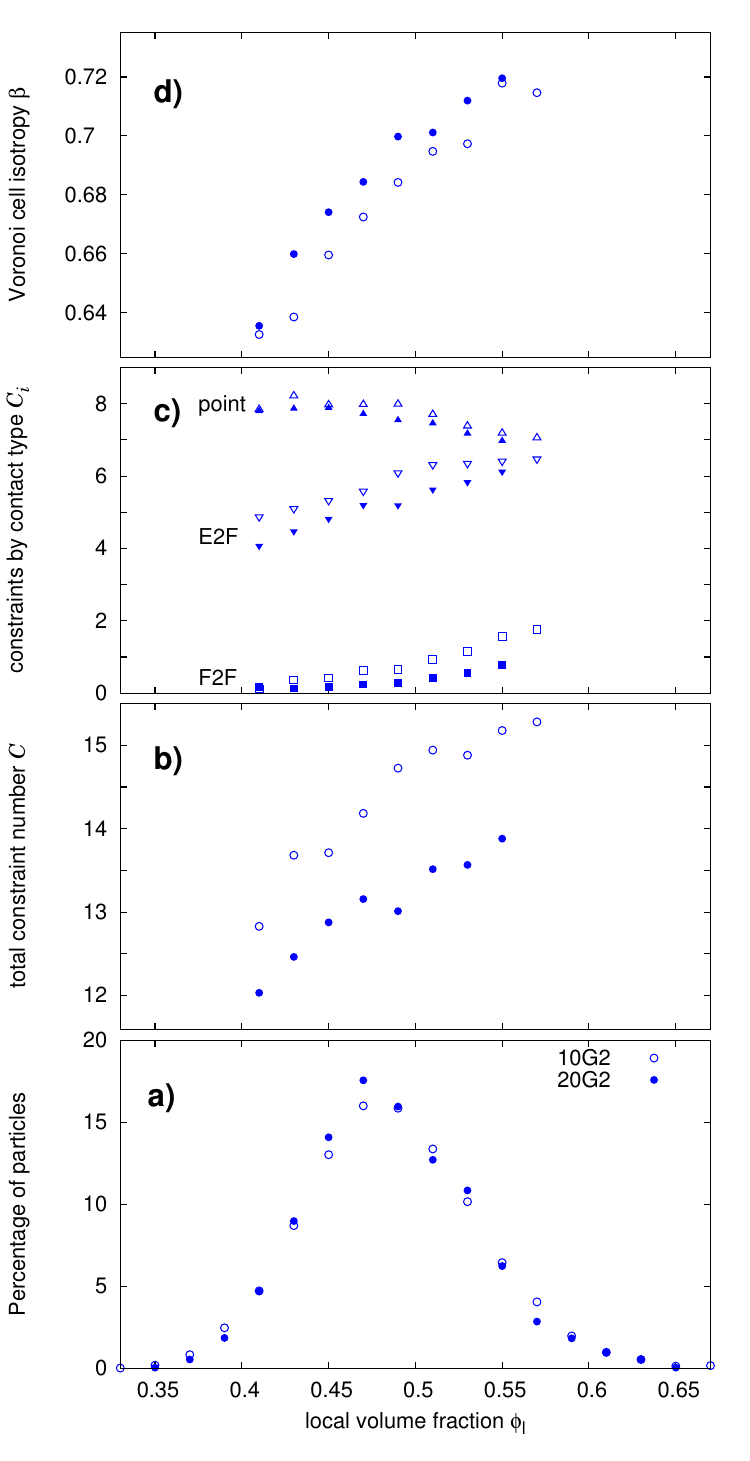}
\caption{Comparing two packings which have been only weakly excited (10 respectively 20 taps at 2g)
after filling of the container.  Their distribution of local volume fractions (a) is identical within experimental errors. 
The total number of constraints (b) is consistently higher for experiment {\tt 10G2}, this increase is due to all
three contact types as shown in panel c). Contrary  to the E2F and F2F contacts, the slope $d C_{point}/ d \phi_\textnormal{local}$ 
is negative. 
Finally the Voronoi cell isotropy (d) is larger for the particles which are less constraint. 
Data points in panels b to d  correspond to all bins which contain more than 3 \% of the totally number of particles.
}
\label{fig:multi_links}
\end{figure}

\begin{figure*}[t]
\includegraphics[width=\textwidth]{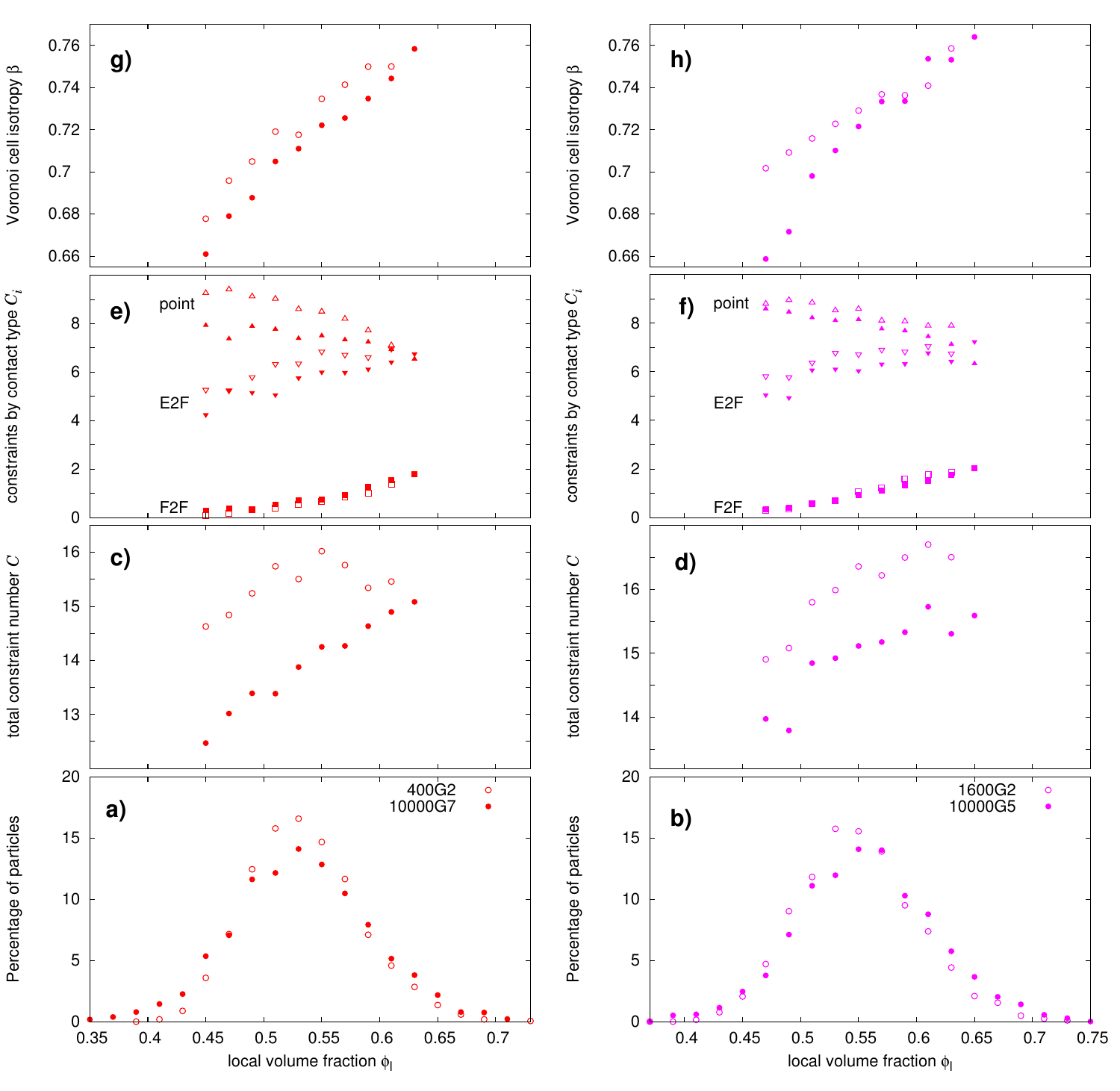}
\caption{Comparing pairs of packings which have been tapped at 2g (open symbols) respectively 5g and 7g (closed symbols). 
For both pairs the distribution of local volume fractions (a) of the higher constraint packing is narrower.
Panels e) and f) show that the increase in total  $C$ (displayed in panels c) and d)) 
originates only from the E2F and point contacts. Similar to figure \ref{fig:multi_links}c
the slope $C_{point}$ with respect to  $\phi_\textnormal{local}$  is negative. However, in contrast to 
figure \ref{fig:multi_links}d, the Voronoi cell isotropy (shown in panels g and h) 
is larger for the more constraint particles. 
Data points in panels c to h  correspond to all bins which contain more than 3 \% of the totally number of particles.
}
\label{fig:multi_merge}
\end{figure*}

As already discussed in the introduction, a global model, i.e.~understanding $C$ as a function of $\phi_\textnormal{global}$,
is  only appropriate for compressible particles where pressure acts as a hidden variable 
controlling both $C$ and  $\phi_\textnormal{global}$ \cite{schroeter:17}. 
Hard and frictional particles posses no mechanism how $\phi_\textnormal{global}$ could
control the number of contacts they form. And even
if we assume the existence of a particle scale demon,  
this demon would be unable to compute $\phi_\textnormal{global}$ by averaging 
the $\phi_\textnormal{local}$ values of the surrounding particles because these values  
are spatially correlated \cite{lechenault:06,zhao_sc:12}.

Because the formation of contacts in granular materials is a local process, we need to describe it  
with locally defined variables. The most important of these is $\phi_\textnormal{local}$:
it has been shown experimentally\cite{aste:06,schaller:15}   that in packings of spheres
the most likely contact number of a particles depends only on $\phi_\textnormal{local}$. 
And {\it not} on $\phi_\textnormal{global}$. 
In a first approximation, this locality of the contact formation also holds true for ellipsoids packings where next to $\phi_\textnormal{local}$ only the 
aspect ratio $\alpha$ of the particles is required to know for predicting the average contact number \cite{schaller:15}.

For the highly anisotropic tetrahedra, it can be expected that both the shape of the free 
volume surrounding them, and their relative orientation with respect to the neighboring
particles might influence the number of contacts they form. 
We will describe the former with the isotropy index $\beta_0^{2,0}$ of the Voronoi cells
and the latter by the contribution of the individual contact types $C_i$ (where we acknowledge
that the latter choice blurs the line between control parameter and result).

The results of our local analysis can be found in figure \ref{fig:multi_links} for the two only marginally shaken packings
and in figure \ref{fig:multi_merge} for the comparison of the different tapping strengths 2 g and 5/7 g.

\subsubsection{Distribution of local volume fractions}
The harmonic mean of $\phi_\textnormal{local}$ (which is how  $\phi_\textnormal{global}$ is defined for packings of monodisperse particles) is roughly the same 
within each pair of tetrahedra packings.  For  spheres \cite{schaller:15}  
one could therefore expect that their $C$ values are also identical.  
A possible explanation for their in fact different values of $C$ would be the following idea:  As in the case of spheres and ellipsoids, there exist some universal, nonlinear functions for $C_i(\phi_\textnormal{local})$. But at the same time the distributions of the $\phi_\textnormal{local}$ values of the two samples are skewed in different directions, resulting in different mean values of $C$.

However, Figures \ref{fig:multi_links}a, \ref{fig:multi_merge}a,  and \ref{fig:multi_merge}b  show that this idea does not apply: 
There is a  small shift corresponding to the 
slightly different mean values and a small change in the width of the distribution (quantified below), but no pronounced asymmetry.

For spheres and ellipsoids it has been shown \cite{schaller:15}
that the standard deviation $\sigma$ of the distribution of the $\phi_\textnormal{local}$ values does only depend on the harmonic mean of $\phi_\textnormal{local}$, not even the aspect ratio of the ellipsoids.
In table \ref{tab:choice} the $\sigma$ of our experiments are listed.
The $\sigma$ value of  {\tt 10G2} is 6.5\% larger than the one of its less constraint partner; which seems to be within experimental errors.
When comparing the pairs of packings created with different accelerations, the less constraint packings have the broader distributions; the $\sigma$  value of  {\tt 10000G7} is  25\%,  the $\sigma$  of  {\tt 10000G5} is 14 \% larger than their
counterparts. We note however, that such a global analysis is susceptible  to spatial gradients within the sample. While for all other samples the fluctuations of $\phi_\textnormal{local}$ as a function of height are smaller than $\pm$ 0.01,  $\phi_\textnormal{local}$ of {\tt 10000G7} changes $\approx$ 0.03. This explains to some extent the larger $\sigma$. In summary, the differences in $\sigma$ seem to be rather small between the samples. 


\subsubsection{Contact probabilities depend on local volume fraction}
\label{sec:local_constraint_contribution}

A more detailed analysis on how $C$ depends on $\phi_\textnormal{local}$ is shown in figures
\ref{fig:multi_links}b, \ref{fig:multi_merge}c, and \ref{fig:multi_merge}d. In all three 
pairs of experiments the values of $C(\phi_\textnormal{local})$ are consistently larger for the 
higher constraint packing. However, the functional form of the $C(\phi_\textnormal{local})$ curves 
is not identical.
$C(\phi_\textnormal{local})$ increases monotonically
for both marginally tapped samples in  figure \ref{fig:multi_links}b; 
the two curves are only shifted vertically against each other.
The situation is more complex for the two pairs tapped with different strengths
(figure \ref{fig:multi_merge}c and d): for the packings {\tt 400G2}  
$C(\phi_\textnormal{local})$ seems to approach a plateau, for the other three packings
$C(\phi_\textnormal{local})$ increases monotonically  but with different slopes.
  
As described in section \ref{sec:tetra_pack}, the total number of constraints  
$C$ equals  $ \sum_i C_i$ where $i$ goes over the three contact types.
Figures \ref{fig:multi_links}c, \ref{fig:multi_merge}e, and \ref{fig:multi_merge}f
display how the contribution of the three different contact types changes with $\phi_\textnormal{local}$.
As already indicated by table \ref{tab:choice}, point contacts
provide the largest contribution to $C$, this is also true for all individual values 
of $\phi_\textnormal{local}$.  Interestingly however, the slope $d C_{point}/ d \phi_\textnormal{local}$
is negative while it is positive for the E2F and F2F contacts, as it is for contacts between spheres or 
ellipsoids \cite{schaller:15}.   
This agrees with the intuitive notion that the closer two tetrahedra get to each other, the more likely it is that
their flat faces or straight edges align with each other and the corresponding contact changes from a point 
contact to a higher constrained one.

For all pairs of packings the difference in preparation is visible in the E2F and
point contacts: in a first approximation the $C_{E2F}(\phi_\textnormal{local})$ and  $C_{point}(\phi_\textnormal{local})$ curves 
are shifted vertically while their average slope is preserved.

\begin{figure}[t]
\centering
\includegraphics[width=0.5\textwidth]{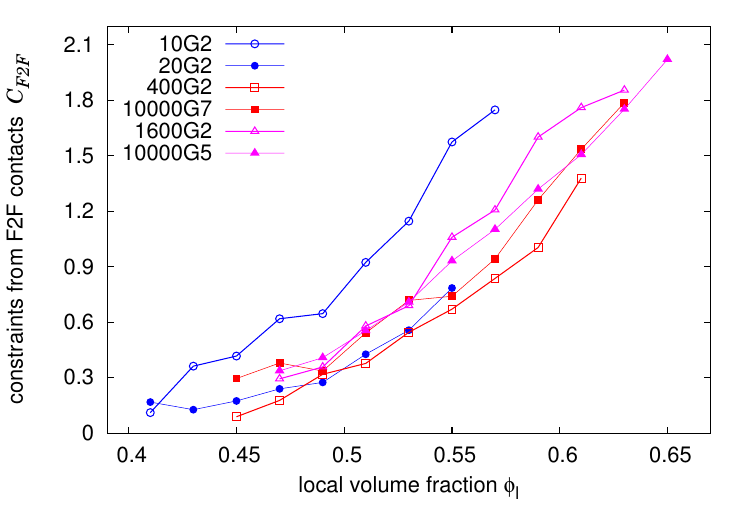}
\caption{Constraints due to F2F contacts seem to fall on a universal curve for samples which have been 
sufficiently tapped (I.e.~all samples except for {\tt 10G2} ). 
Included are all bins in $\phi_\textnormal{local}$ which amount to at least 3 \% of the total number of particles.
}
\label{fig:F2F_all}
\end{figure}

The situation is different for the F2F contacts:
for the pairs tapped at different accelerations the $C_{F2F}(\phi_\textnormal{local})$ curves
fall on top of each other. Figure \ref{fig:F2F_all} demonstrates
that this agreement goes even further: with the exception of the loose {\tt 10G2}  sample all other  
$C_{F2F}(\phi_\textnormal{local})$ curves coincide within experimental errors. 
Which points to the existence of a preparation-independent master
function controlling the number of F2F contacts in samples which have been sufficiently tapped. 
  
Some motivation for this different behavior of the  F2F contacts can be derived from the high value
of their constraint multiplier $m_{F2F}$: Once a particle has formed an F2F contact with another particle, 
all three rotational degrees of freedom are completely blocked for both particles. 
Any additional F2F contact (i.e.~increase in  $C_{F2F}$)
depends therefore solely on the capability of another incoming particle to align 
itself such that there is a 180 degree angle between the face normal vectors. The probability that 
such a rotation is possible will depend strongly on the available space (i.e.~$\phi_\textnormal{local}$). 


In contrast, pairs of particles which have established an E2F or point contact between themselves, retain one or even three 
rotational degrees of freedom to facilitate another contact. Which means that  geometrical factors, 
other than $\phi_\textnormal{local}$, will play a role in determining the probability if another contact of the same type can 
be formed.

\subsubsection{Voronoi cell isotropy}
\label{sec:isotropy_local}

Figures \ref{fig:multi_links}d, \ref{fig:multi_merge}g, and \ref{fig:multi_merge}h display the monotonic  
increase of the  isotropy  index $\beta_0^{2,0}$ with  $\phi_\textnormal{local}$ in all our experiments. 
While such an increase is in agreement with the results for  sphere packings \cite{schroeder-turk:10}, 
the fact that there is no universal curve  for $\beta_0^{2,0} (\phi_\textnormal{local})$ differs
from the behavior of spheres but is compatible with packings of ellipsoids \cite{schaller:15b}.

Moreover, the $\beta_0^{2,0} (\phi_\textnormal{local})$ curves reinforce the result already shown in table \ref{tab:choice}:
for the only marginally tapped samples the lower constraint packing contains the more isotropic
Voronoi cells, for the samples tapped at different levels of g, the higher constraint packings are more isotropic.
This signifies that $\beta_0^{2,0}$ is {\it not} likely to be the hidden parameter connecting the 
preparation history with the constraint number of the packing.  
 
\subsection{Influence of electrostatics?}
A possible candidate for the physical parameter controlling the differences between the packings shaken at different accelerations are electrostatic interactions between the particles. It has recently been shown \cite{schella:17} that electric charges change the contact probabilities in bidisperse sphere packings. For tetrahedra we did observe that in the experiments at 5g and 7g more (polypropylene) tetrahedra did stick to the (PMMA) sidewalls of the shaking container than in the experiments at 2g. 
And because mixing will distribute charges inside the sample, they would also qualify as a locally defined parameter. However, we could expect that such charges would influence the formation of F2F contacts most (due to their extended proximity between the surfaces).
In contrast to this argument, figure \ref{fig:F2F_all} shows that the F2F contacts are the least involved in the differences between the packings.

\section{Conclusions}

Because tetrahedra have four flat faces, their packings do not only posses point contacts similar to 
sphere packings but also edge-to-face and face-to-face contacts which fix larger numbers of geometrical
constraints than point contacts. The number of point and edge-to-face contacts a given tetrahedron forms can not be 
predicted from the global or local volume fraction; the preparation history seems to be encoded in 
still another parameter.
In contrast, the formation of face-to-face contacts in sufficiently tapped samples seems to be a function of the local volume fraction alone, increasing with increasing proximity between particles.  

While the isotropy of the Voronoi volume shape,
and to some extent also the width of the local volume fraction distribution, do  depend on the preparation history of the packing, neither of these two parameters qualifies as 
the missing control parameter. 
Our results constitute therefore solely a phenomenological description of the history-dependence 
in tetrahedra packings. While they neither support nor rule out the feasibility of a mean-field approach, they imply 
that the jamming paradigm in its present form will not be an adequate description of the contact formation. 
More work is needed.

\begin{acknowledgments}
We acknowledge experimental support by Wolf Keiderling.
\end{acknowledgments}


\begin{thebibliography}{67}
\bibitem [{not()}]{note_friction}%
  \BibitemOpen
  \href@noop {} {}\bibinfo {note} {$\mu_s$ is computed from the tangens of the
  inclination angle where a face to face contact starts to slide.}\BibitemShut
  {Stop}%
\bibitem [{zen()}]{zenodo}%
  \BibitemOpen
  \href@noop {} {}\bibinfo {note} {DOI from Zenodo; data upload is only
  possible after paper is accepted by a journal.}\BibitemShut {Stop}%
\end{thebibliography}
\end{document}